\documentclass[12pt,preprint]{aastex}

\usepackage{graphicx}
\usepackage{psfig}
\usepackage{natbib}
\usepackage{amsfonts}
\usepackage{amssymb}
\usepackage{amsmath}
\usepackage{indentfirst}

\newcommand{\ud}{\mathrm{d}}

\shorttitle{High Energy Emission from GRBs}
\shortauthors{Gialis \& Pelletier}

\begin{document}


\title{High Energy Emission and Cosmic Rays from Gamma-Ray Bursts}

\author{D. Gialis and G. Pelletier\altaffilmark{1}}
\affil{Laboratoire d'Astrophysique de Grenoble,\\ Universit\'e Joseph Fourier, Grenoble, F-38041, France}
\email{Denis.Gialis@obs.ujf-grenoble.fr}
\email{Guy.Pelletier@obs.ujf-grenoble.fr}

\altaffiltext{1}{Institut Universitaire de France.}

\begin{abstract}
The paper is devoted to the analysis of particle acceleration in Gamma-Ray Bursts (GRBs) and its radiative consequences. Therefore we get on one hand constraints on the physics and on the other hand possible signatures of particle acceleration that could be recorded by the new gamma ray instruments. In a recent paper we have shown that UHECRs can be generated in GRBs even with conservative assumptions on the magnetic field and the scattering capability of its perturbations, provided that a suitable relativistic Fermi process is at work during the so-called "internal shock" phase. In this paper we extend the analysis of the consequences of these assumptions to the whole prompt emission of both electrons and protons. Indeed, assuming that the magnetic field decays in $1/r^2$ and that the scattering time of particles is longer than the Bohm's assumption, in particular with a rule derived from Kolmogorov scaling, we show that the five following events naturally happen with no other parameter adaptation that the intensity of the magnetic field, that turns out to be subequipartition:  i) UHECRs can be generated with a sufficient flux ($\simeq$ 1 km$^{-2}$yr$^{-1}$) within the GZK-sphere to account for the CR-spectrum at the ankle (in the previous paper, we showed that the associated p$\gamma$-neutrino emission is tiny). ii) The peak energy of the gamma spectrum around $100$ keV, namely the so-called $E_{peak}$, is conveniently explained. iii) A thermal component below the $E_{peak}$ is often unavoidable. iv) The cosmic rays could radiate gamma rays around $67$ MeV (in the co-moving frame, which implies $\simeq 20$ GeV for the observer) due to $\pi^0$-decay and a low energy neutrino emission (around 0.2 GeV) associated to neutron decay and also  neutrinos of energy between 5 and 150 GeV from muon decay (as predicted in the previous paper). v) The UHECRs radiate high energy gamma rays between a few hundreds of MeV and $10$ GeV (taking the pair creation process into account) due to their synchrotron emission with a sufficient flux to be observable.

\end{abstract}


\section{Introduction}

Whereas the afterglow emission of Gamma-Ray Bursts (hereafter GRBs) has been well investigated and  provided a strong support to the "fireball model" \citep{ReesMeszaros92}, the origin of the prompt emission of is not well established yet. The most widely accepted scenario explaining the burst emission is the internal shock model \citep{ReesMeszaros94}: according to this model, the prompt gamma-ray emission results from either the synchrotron emission of accelerated electrons or the inverse Compton scattering off the synchrotron photons, the relativistic electrons being produced by the collision of relativistic shells. But, even if the prompt emission spectrum is correctly fitted by the empirical Band Function \citep{Band1993}, no physical process satisfactorily explains it especially below the peak energy ($E_{peak}$). More recently, the possibility of a thermal component as the low energy part ($\lesssim 100$ keV) of the prompt emission spectrum had been studied \citep{Ghi03} and successfully compared to observations. In previous papers (\citet{GialisPelletier03}, \citet{DGGP04}), we were interested by the issue of the particle acceleration in the internal shocks and by the resulting cosmic ray generation: our results significantly changed the usual interpretation of the Fermi acceleration \citep{Waxman95} and this led to propose an additionnal Fermi process to reach the goal of the UHECR production. In this paper, we intend to emphasize the consequences of the particle acceleration on high energy gamma-ray emission and we give an interpretation of the energy spectrum of the prompt emission. We also predict both the production of lower energy photons in the UV-X-ray range and a very high energy emission (between a few hundreds of MeV and $10$ GeV, free from opacity effect) which should be detectable, for instance, with the GLAST experiment.   \\
The paper is organized as follows: in the section \ref{sec1}, we briefly describe the fireball dynamics from the central object to the deceleration radius. We also calculate the photospheric radius and we determine some radiative parameters we need for this paper. The section \ref{PartAcc} summarizes our previous results \citep{GialisPelletier03} concerning the cosmic ray acceleration in GRBs and extends the study to the electron population. We analyse the consequences of the particle acceleration on high energy emission in the section \ref{HEE}. The last section of this paper is our conclusion on the new results that we have obtained and which could be confirmed by forthcoming experiments such as HESS2 and GLAST.

\section{\textsf{Dynamics and radiative parameters}}
\label{sec1}

\subsection{\textsf{Outline of the fireball dynamics}}
In this subsection, we briefly summarize all the results we need for this
paper that describe the fireball dynamics \citep{Meszaros93} before the deceleration radius where an external shock starts.\\

We choose to describe the outflow with some simplified hypothesis: the wind flow can be considered to be a set of discrete shells which are successively emitted with an energy $E_{s}=E/N_{s}$, where $N_{s}$ is the total number of shells and $E$ the total energy released by the fireball (\citet{Daigne98}). We assume that the total energy radiated in gamma-rays $E_{\gamma}$ is a sizeable fraction of $E$. According to the observations, this flow is collimated with an average opening solid angle, $\Omega$, of about $4\pi/500$ (\citet{Frail2001}). The wind flow duration, namely $\Delta t_{w}$,
provides with an interval of shell number which is  $1 \leq N_{s} \leq
c\, \Delta t_{w}/r_{0}$ where $r_{0}\simeq 10^{7}$ cm is the size of the central object. In a primeval stage of a shell expansion, the radiative pressure gives the temperature, $T$, of the completely optical thick plasma which is mainly composed by electron-positron pairs and by electrons and protons beyond the pair annihilation radius. This temperature can initially be defined by
\begin{eqnarray}
T=\left( \frac{3\,E_{s}}{4\pi\, a_{s}\, r_{0}^{3}} \right)^{1/4}\,,
\end{eqnarray}
where $E_{s}$ is the energy of an emitted shell and $a_{s}=7.56\times 10^{-15}$(c.g.s) is the Stefan constant.
For an energy $E_{s}$ of the order of $10^{51}$ erg, the temperature at $r_{0}$ is about a few MeV.\\
A characteristic baryon loading parameter, $\eta$, is defined as the ratio between $E$ and the baryon rest mass energy :
\begin{eqnarray}
\eta  = \frac{E}{M_{b}\, c^{2}} \gg 1
\end{eqnarray}
where $M_{b}$ is the total baryonic mass ejected. The value of the baryon loading parameter is usually considered between $10^{2}$ and $10^{3}$, in order to solve the ``compactness problem''. \\
In a first stage, the ejected shell follows an adiabatic expansion in the surrounding medium and its internal energy is progressively converted into kinetic energy. Hereafter, we define the stationary frame as the rest frame of the central object. In the stationary frame, we can define a radius, $r_{s}$, where the
kinetic energy of baryonic matter reaches its saturation value. At
this moment, the Lorentz factor $\Gamma$ of a given shell is close to $\eta$ which constitutes an average value.  This last point is important to understand the internal shock model.\\
In the same frame, the shell thickness, $\Delta r$,  remains constant and equal to $r_{0}$ until the broadening radius $r_{b} > r_{s}$ (\citet{Goodman86}, \citet{Meszaros93}). Beyond this radius, the shell thickness becomes $\Delta r \simeq r/2\Gamma^{2}\simeq r/\eta^{2}$, so that the broadening radius is about $\eta^{2}\,r_{0}$.\\
In the co-moving frame of a shell, entropy and energy conservation before $r_{s}$ give the evolution of the Lorentz factor $\Gamma$, the temperature $T$ and the shell thickness, namely $\Delta R = \Gamma \,\Delta r$: we have $\Gamma \propto r$, $T \propto r^{-1}$ and $\Delta R = r/r_{0}$. Thus, the saturation radius $r_{s}$ is equal to $\eta\,r_{0}$.\\
Beyond $r_{s}$, the Lorentz factor $\Gamma$ remains constant and the previous evolution laws become $\Gamma \simeq \eta$ and $T \propto r^{-2/3}$. We can also write
\begin{equation}
T\simeq 17 \times \left( \frac{T(r_{0})}{5\, MeV} \right) \left(
\frac{\eta}{300} \right)^{-1} \left( \frac{r}{r_{s}} \right)^{-2/3}\, \,
keV
\label{temp}
\end{equation}
At last, the shell thickness in the co-moving frame is such that $\Delta R = \eta\,r_{0}$ before $r_{b}$ and $\Delta R = r/\eta$ beyond. \\
The internal shock model has been designed  \citep{Meszaros93} in order to account for rapid variability observed in GRB light curves and which can reach the millisecond. The internal shock model scheme is the following one: let us consider two shells leaving the central engine separated by a time interval $\Delta t$, respectively with the Lorentz factors $\Gamma_{1}$ and $\Gamma_{2}$ such that $\Gamma_{2}>\Gamma_{1}$. A collision occurs at the date:
\begin{eqnarray}
t_{c}\simeq \frac{2\Gamma_{1}^{2}\,\Gamma_{2}^{2}}{\Gamma_{2}^{2}-\Gamma_{1}^{2}}\, \Delta t\,.
\end{eqnarray}
Assuming an instantaneous shock pulse, some time spreading, $\Delta t_s=t_c/2\Gamma^2$, is observed. Thus, the shortest variabilities which are observed ($\sim$ 1 ms) will be such that $t_{b}/2\Gamma^2\simeq r_{0}/2c$. These correspond to typical time scale associated to the size of a black hole of a few tens solar masses (namely $r_{0}/c$). Such a first collision takes place around the distance $r_{b}$. Longer variations correspond to collisions at a more remote distance until a maximum distance determined by the duration of the flow $\Delta t_w$. This maximum distance is $r_{max} \sim r_{b}\, c\, \Delta t_{w}/r_{0}$, with $c\, \Delta t_{w}/r_{0} \simeq 3\times 10^3(\Delta t_w/1s)$, which gives a proper length of the flow in the co-moving frame $\ell_{0}=\beta\,c\,\Gamma\,\Delta t_{w}$. The duration of the flow during the internal shock phase is therefore $\Delta t_{max}\sim (r_{b}/r_{0})\,\Delta t_{w} \sim \eta^{2}\,\Delta t_{w}$. This phenomena is observed during a time interval shortened by the propagation effect, namely $\Delta t_{obs}=(1-\beta)\, \Delta t_{max} \simeq \Delta t_{max}/2\eta^{2} \sim \Delta t_{w}$. The previous value of $r_{max}$  is not far from the deceleration radius, $r_{d}$, of the shells which is about $10^{16}$cm. Also, the Fermi acceleration of particles, which is usually considered \citep{Waxman95}, takes place in the range that extends from $r_b$ to $r_{d}$, namely the internal shock phase.

\subsection{\textsf{Radiative parameters}}
\label{radpar}

We have seen that, in a primeval stage, the ejected plasma is optically thick with respect to the Compton scattering. Using the results of the previous subsection, we propose here to determine the photospheric radius and we will define some radiative parameters .  \\

First, it can easily be checked that a typical shell width $\Delta R$ becomes smaller than the flow transverse radius after a short while, when $r> \eta
\sqrt{\pi/4\Omega}\,r_{0}$ which is comparable to $r_{s}$. It will
turn out that the photosphere is located at a much larger distance for
large enough $\eta$  and therefore the opacity of a shell is determined by its width. Assuming the temperature is such that  $\bar{\gamma}_{e}\,h\nu \ll m_{e}\,c^2$ where $\bar{\gamma}_{e}$ is the average electron Lorentz factor ($\bar{\gamma}_{e}\simeq 1$ beyond the pair annihilation radius), the optical depth can be defined by  $\tau_{\star} = \sigma_{T}\, n_{e}\,\Delta R$ with $\sigma_{T}$ is the Thomson cross section. Because of the plasma neutrality, we have $n_{e}\simeq n_{p}$, so that the co-moving electron density can be written
\begin{equation}
n_{e} = \frac{\xi_{s}\,E}{\Gamma\, \Omega\, r^{2}\, \Delta R\,m_{p}\,c^{2}}\,,
\label{eq:NB}
\end{equation}
where $\xi_{s}=1/N_{s}$. Thus, the optical depth is
\begin{equation}
        \tau_{\star} = \frac{\xi_{s}\,\sigma_{T}\,E}{\Omega\,r^{2}
        \,m_{p}\,c^{2}\,\eta} \, .
        \label{TAUS}
\end{equation}
We can define a critical value for $\eta$ such that the photospheric
radius ($\tau_{\star}=1$) is located at $r_{b}$, where shock acceleration starts. This critical value $\eta_{\star}$ is given by
\begin{equation}
\eta_{\star} \simeq 1780\,\left(\frac{\xi_{s}}{10^{-1}}\right)^{1/5}\left(\frac{\Omega/4\pi}{2\times 10^{-3}}\right)^{-1/5}\,\left(\frac{E}{10^{51}\,erg}\right)^{1/5}\,.
\label{ETAS}
\end{equation}
For a GRB with $N_{s}\simeq 100$, $\eta_{\star}\simeq 1100$ and drops around 450 for a long GRB with about $10^{4}$ shells.\\
\noindent Thus, we can express the photospheric radius $r_{\star}$ as
\begin{eqnarray}
r_{\star} = r_{b}\,\left(\frac{\eta_{\star}}{\eta}\right)^{5/2}\,.
\label{rstar}
\end{eqnarray}
For usual values of $\eta$ and according to Eq. (\ref{ETAS}), we conclude that the photospheric radius can be over $r_{b}$, so that the internal shocks start accelerating particles in an optically thick plasma. We will analyse some consequences in the next sections. Also, beyond $r_{\star}$, one can consider that photons and electrons decouple and, if $r>r_{b}$, electrons can be accelerated via the Fermi acceleration in the internal shocks.\\

\begin{figure}[!h]
\begin{center}
\includegraphics[totalheight=6.5cm]{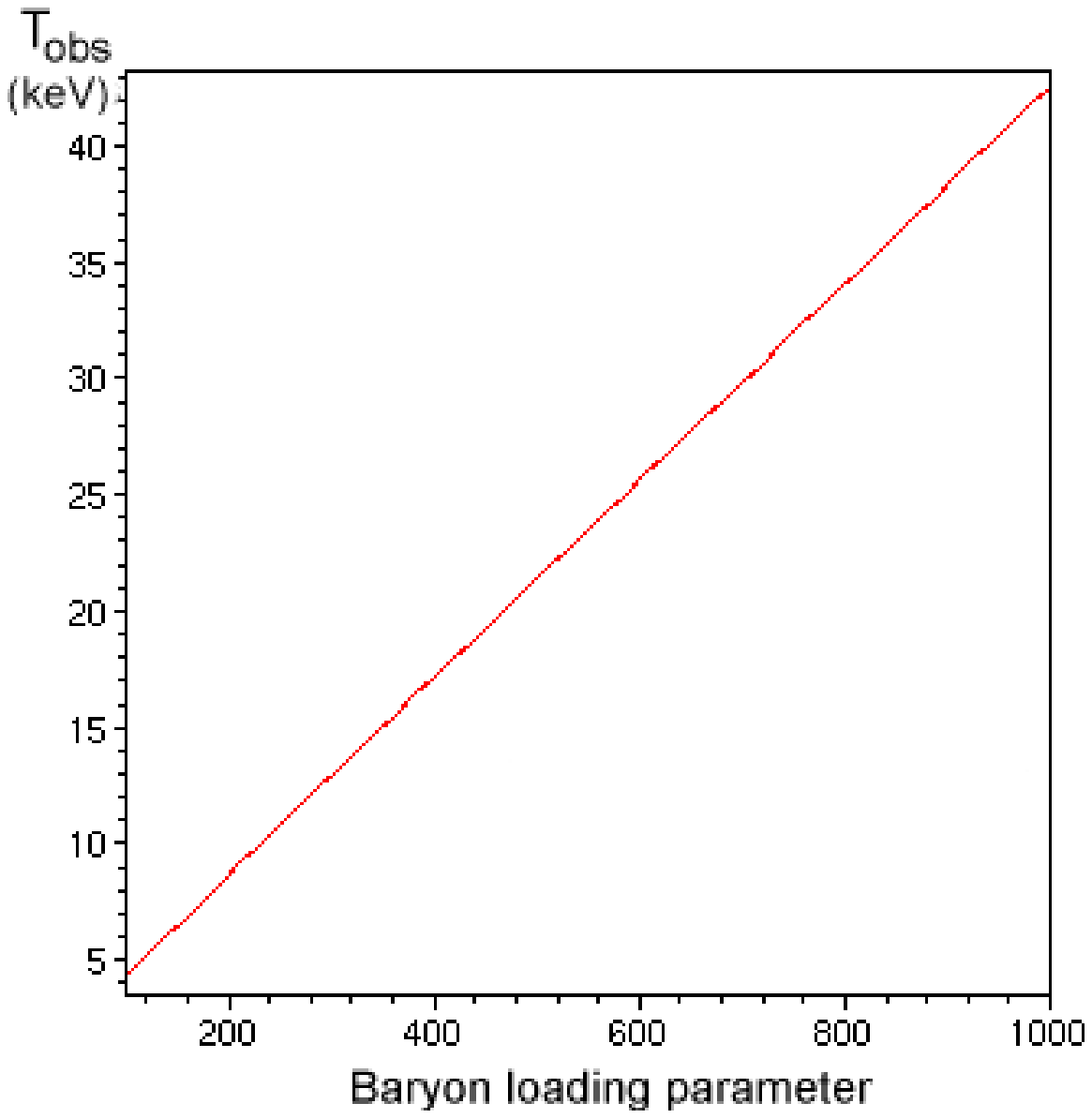}
\includegraphics[totalheight=6.5cm]{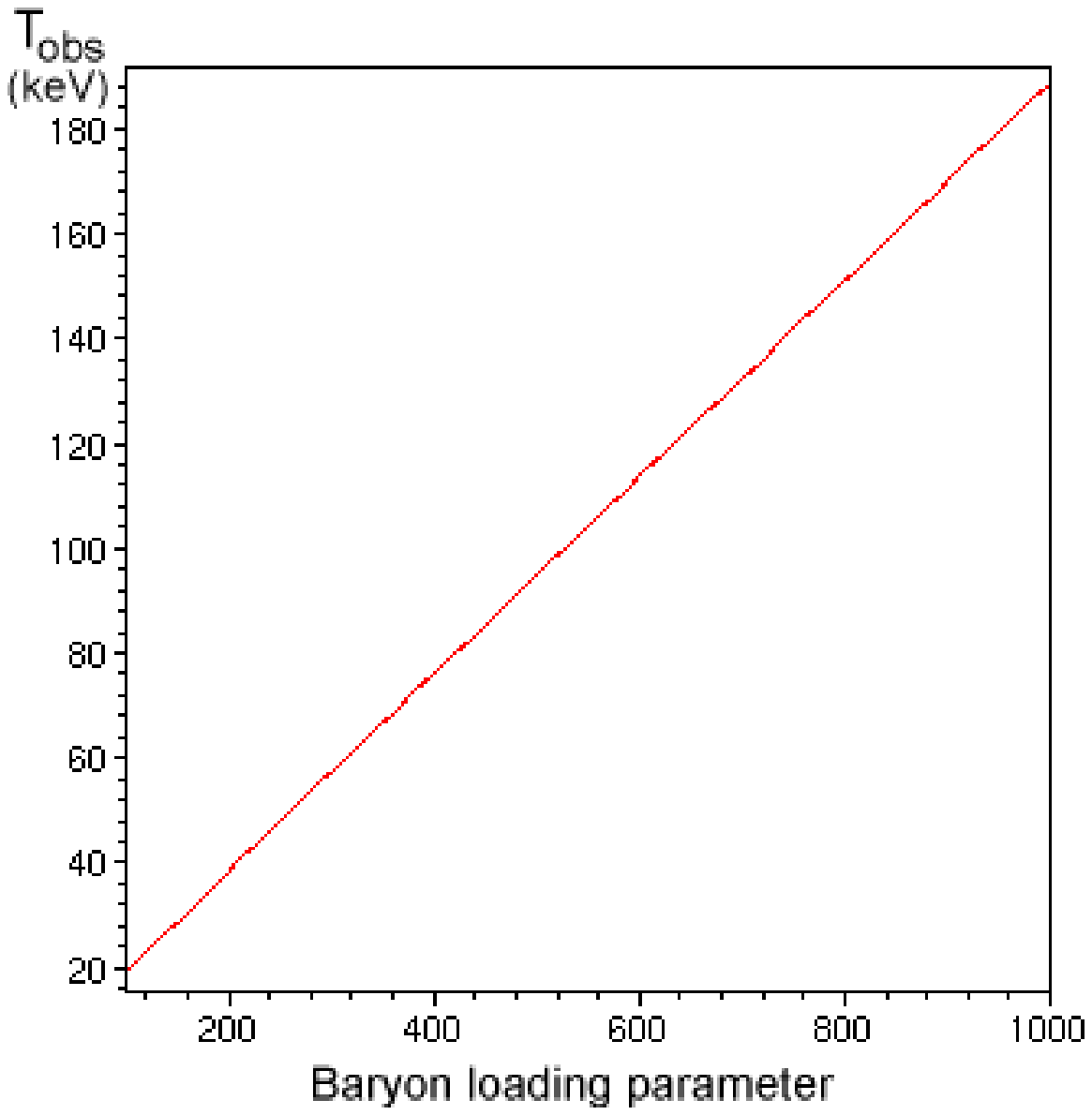}\\
\includegraphics[totalheight=6.5cm]{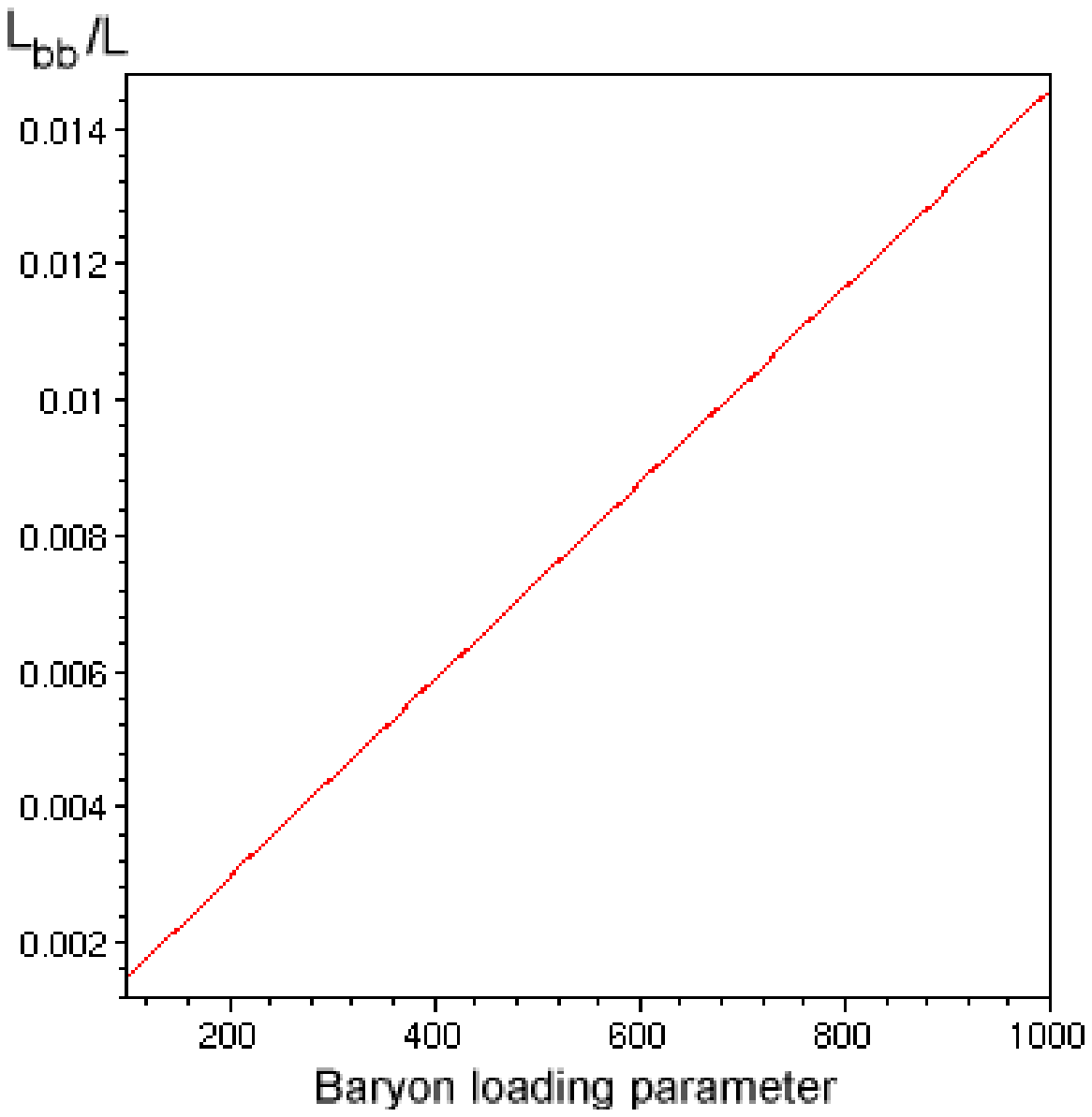}
\includegraphics[totalheight=6.5cm]{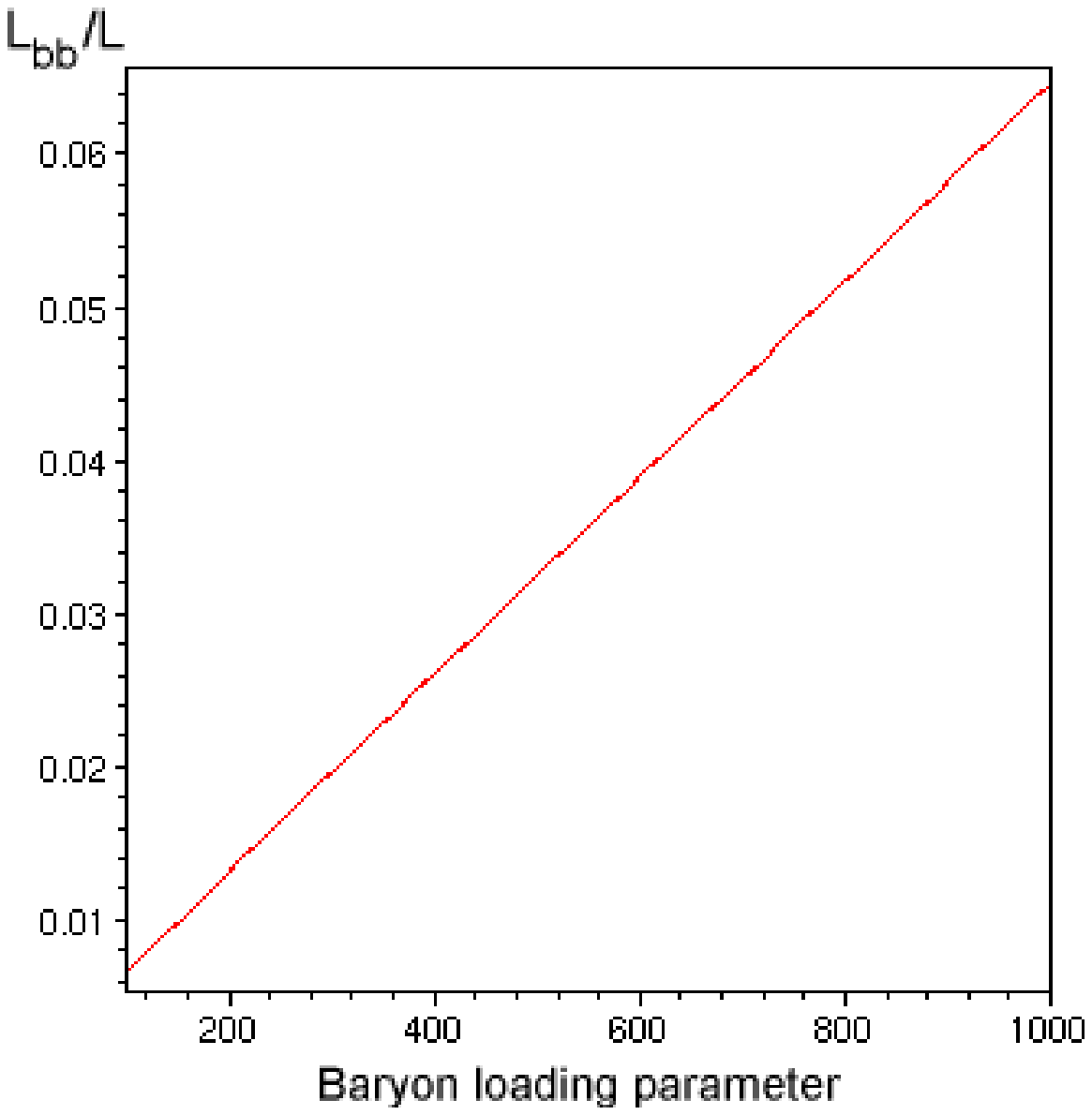}
\caption[10pts]{Top: Variation of the black body radiation temperature, measured in the observer frame ($T_{obs}\simeq \eta\,T_{\star}$) as a function of the baryon loading parameter $\eta$. Bottom: Variation of the ratio $L_{bb}/L_{\gamma}$.  Left: the GRBs with $N_{s}=100$ and $\eta_{\star}=1100$. Right: the long GRBs with $N_{s}=10^{4}$ and $\eta_{\star}=450$.}
\label{therm1}
\end{center}
\end{figure}

\noindent Considering the resulting black body emission at the photospheric radius, Eq. (\ref{temp}) gives the temperature which is such that :
\begin{equation}
T_{\star}=0.37\,\left( \frac{T(r_{0})}{5\, MeV} \right)\left(\frac{\eta}{300}\right)^{-5/3}\left(\frac{\eta_{\star}}{\eta}\right)^{-5/3}\,keV.
\label{Ts1}
\end{equation}
This result must be compared to the following one: we have $\chi\,L_{\gamma}=\eta^{2}\,\Omega\,r_{\star}^{2}\,\sigma\,T_{\star}^{4}$ where $\chi$ is defined as the ratio between the average black body luminosity, namely $L_{bb}$, and the GRB gamma-ray luminosity $L_{\gamma}\lesssim E/\Delta t_{w}$. We deduce the expression
\begin{eqnarray}
T_{\star}=0.47\,\left( \frac{\chi}{10^{-1}} \right)^{1/4}
\left(\frac{\Omega/4\pi}{2\times 10^{-3}}\right)^{-1/4} \left(\frac{L_{\gamma}}{10^{51}\,erg.s^{-1}}\right)^{1/4}
\left(\frac{\eta}{300}\right)^{-3/2}\left(\frac{\eta_{\star}}{\eta}\right)^{-5/4}\,keV.
\label{Ts2}
\end{eqnarray}
The comparison between Eqs. (\ref{Ts1}) and (\ref{Ts2}) leads to a ratio $L_{bb}/L_{\gamma}$ easily reaching a few percent:
\begin{equation}
\frac{L_{bb}}{L_{\gamma}}=3.8\times10^{-2}\left(\frac{T(r_{0})}{5\,MeV}\right)^{4}
\left(\frac{\Omega/4\pi}{2\times 10^{-3}}\right)
\left(\frac{L_{\gamma}}{10^{51}\,erg.s^{-1}}\right)^{-1}
\left(\frac{\eta}{300}\right)^{-2/3}\left(\frac{\eta_{\star}}{\eta}\right)^{-5/3}\,.
\label{khi}
\end{equation}
Thus, the result is a thermal component in the GRB spectrum which can be observed, before an higher energy emission, in the range 10 keV $-$ 180 keV in the observer frame as we see in Fig. \ref{therm1}. Some authors have already considered this possibility (see e.g \citet{Ghi03}) which seems to be consistent with observations. Moreover, we note that Eqs. (\ref{Ts1}) and (\ref{khi}) indicate a more important thermal component around 100 keV (with $\eta=400$) for a high number of shells because of a small $\eta_{\star}$ parameter.

\section{\textsf{Particle acceleration in the internal shock phase}}
\label{PartAcc}

\subsection{\textsf{Proton acceleration and cosmic ray generation}}
\label{protacc}

The Fermi acceleration (first or second order) in the internal shock model is usually considered \citep{Waxman95} as mildly or sub-relativistic with a characteristic time proportional to the Larmor time (Bohm scaling). However, in a previous paper \citep{GialisPelletier03}, we have shown that this assumption is not realistic regarding the magnetic energy depletion time. Moreover, the Fermi acceleration time depends on the mean free path, $\bar{\ell}$, of the particle in an irregular magnetic field. This length depends on two other lengths, namely the Larmor radius, $r_{L}$, and the correlation length, $\ell_{c}$ : for a turbulence spectrum of magnetic perturbations in a power law of index $\beta$, the following law, which is known in weak turbulence theory, has been extended in the regime of strong turbulence and large rigidities such that $r_{L} < \ell_{c}$ \citep{Casse01}:
\begin{eqnarray}
\bar{\ell}=\frac{r_{L}}{\eta_t}\,\left(\frac{r_{L}}{\ell_{c}}  \right)^{1-\beta}\,,
\end{eqnarray}
where $\eta_t =\frac{<\delta B^{2}>}{<B^2>}$.  The Bohm scaling $\bar{\ell} \sim r_{L}$, which holds for electrostatic turbulence, does not apply with purely magnetic irregularities on large scale; no theory nor numerical simulation has confirmed Bohm's conjecture. The Bohm estimate corresponds only to the specific case where the magnetic field is totally disorganized and the Larmor radius as large as the correlation length which is not the case in GRBs. The realistic situation is the following one: assuming that the correlation length is of the order of a shell thickness, namely $\Delta R$, the characteristic acceleration time is $t_{acc}=\kappa\,t_{L}$ where $t_{L}$ is the Larmor time and $\kappa \propto (r_{L}/\Delta R)^{1-\beta}$.
According to a Kolmogorov scaling with $\beta =5/3$, and defining $\kappa_0$ as the ratio of the acceleration time over the Larmor time for a Larmor radius that equals the correlation length of the magnetic field, we have:
\begin{equation}
t_{acc}\simeq 4.3\times 10^{-3}\left(\frac{\kappa_{0}}{10}\right)\left(\frac{\eta}{300}\right)^{2/3}\left(\frac{B(r_{b})}{10^{4}\,G}\right)^{-1/3}\left(\frac{\epsilon}{1\,GeV}\right)^{1/3}\left(\frac{r}{r_{b}}\right)^{\frac{2+\alpha}{3}}\,s,
\end{equation}
where $\epsilon$ is the energy of a proton and assuming that the magnetic field strength decreases like $r^{-\alpha}$. Although unproved, this scaling is more reasonable and this conservative assumption will lead to sensible results.\\
Comparing this time with the expansion time, $t_{exp}=r/c\eta$, we have shown \citep{GialisPelletier03} that GRBs are unable to produce UHECRs with this acceleration process because of a strong expansion limitation in energy beyond $r_{b}$ which is
\begin{eqnarray}
     \epsilon_{exp} \simeq 1.3\times 10^{4}\left(\frac{\kappa_{0}}{10}\right)^{-3}\left(\frac{\eta}{300}\right)
     \left(\frac{B_{\star}(r_{b})}{10^{4}G}\right)
     \left(\frac{r}{r_{b}}\right)^{1-\alpha}\, GeV \,,
     \label{EEXP}
\end{eqnarray}
This limitation, measured in the co-moving frame, is more severe than the synchrotron one and suggests that we have to consider another type of process to achieve high energy. In a recent paper \citep{DGGP04}, we have proposed a Fermi acceleration process resulting from scattering off relativistic hydromagnetic fronts at the very beginning of the internal shock phase. The efficiency of this process, as shown by numerical simulation, is sufficient to generate a sizeable fraction of UHE cosmic rays. Also, we found that this scenario could constitute a very interesting additionnal acceleration process which stretch the cosmic ray distribution tail obtained with the usual Fermi acceleration process. Moreover, we showed that, for a magnetic field strength decreasing like $r^{-2}$, the high energy cosmic ray population is such that:
\begin{equation}
\frac{\ud N_{\star}}{\ud \gamma}\propto \gamma^{-2}\,,
\label{UHECR1}
\end{equation}
where $N_{\star}$ is the number of UHE cosmic rays and $\gamma$ their Lorentz factor, this spectrum extending over 4 decades from $10^{7}$ to $10^{11}$ GeV in the observer frame.\\
Assuming such a spectrum for the whole proton population in the co-moving Lorentz factor range $[1,10^{9}]$, we will have:\begin{equation}
\frac{\ud N_{\star}}{\ud \gamma}\simeq N_{p}\,\gamma^{-2}\,,
\label{UHECR2}
\end{equation}
where $N_{p}$ is the total number of protons released by GRB. For $N_{p}\simeq 10^{51}$, the number of UHECRs ($\gamma\geq 10^{8}$) generated by GRB is about $10^{43}$. Considering a GRB rate of about 1 per $10^{6}$ Mpc$^{-3}$yr$^{-1}$ (see e.g. \citet{VanPutten03} or \citet{Frail2001}), we deduce that, in the GZK-sphere of $\sim 1$ Gpc$^{3}$, GRBs release $10^{46}$ UHECRs yr$^{-1}$. Because of the intergalactic magnetic field, this UHECR population is almost isotropized, so that we can observe $10^{46}/4\pi\,(1\,Gpc)^{2}\sim 1$ UHECR km$^{-2}$yr$^{-1}$ which is about the observed flux at the ankle in the UHECR spectrum. For a better estimate of this flux, we have to consider more accurately the magnetic field structure in the GZK-sphere (see e.g \citet{Lemoine03} or \citet{Sigl99}).\\  

An interesting consequence of the proton acceleration appears in considering the $pp$-collisions. Indeed, because $n_{p}\simeq n_{e}$, the opacity of $pp$-collisions in a shell, namely $\tau_{pp}=n_{p}\,\sigma_{pp}\,\Delta R$, is such that $\tau_{pp}=(\sigma_{pp}/\sigma_{T})\,\tau_{\star}$. For this reason, we can easily write the radius $r_{pp}$ beyond which there is no more $pp$-collision:
\begin{eqnarray}
r_{pp} = r_{\star}\,\sqrt{\sigma_{pp}/\sigma_{T}} \simeq 0.20\,r_{\star}\,,
\end{eqnarray}
and $r_{pp}>r_{b}$ for $\eta<\eta_{\star}/3$ which is quite possible according to the value of $\eta_{\star}$ calculated in the previous section (this result slightly differs from which we found in \citet{GialisPelletier03} because of a better estimate of $\tau_{\star}$). In this case, the proton acceleration starts in an opaque stage for $pp$-collisions: even if the energy limitation due to $pp$-collisions implies a cut-off energy around 1 GeV for the proton population, there is a possibility of a low energy emission of neutrinos between 5 GeV and 150 GeV which are produced via the $\pi^{+}$ and the $\pi^{-}$ decays \citep{DGGP04}. Moreover, the $pp$-process generates $\pi^{0}$-mesons that decay in photons with an energy of 67 MeV in the co-moving frame. Because we have to consider the Klein-Nishina regime, the cross section, namely $\sigma_{KN}$, is significantly lower than the Thomson cross section ($\sigma_{KN}/\sigma_{T}\simeq 1.7\times 10^{-2}$) so that the associated photospheric radius for these photons is about $0.13\,r_{\star}$. The photons produced at 67 MeV in a thin layer between $0.13\,r_{\star}$ and $r_{pp}$ can cross over the medium without any electronic interaction; this is achieved if $\eta<\eta_{\star}/4$. If one considers the pair creation process for 67 MeV photons on themselves, the cross section is such that $\sigma_{\gamma\gamma}/\sigma_{T}\simeq 2.0\times 10^{-4}$ which leads to a transparency radius $r_{\gamma\gamma}$ of about $10^{-2}\,r_{\star}$ only. At last, these photons cannot interact with the thermal photons ($<0.5$ keV) because the threshold energy is about 7.8 keV.\\

Thus, we predict a possible signature of a such process which could be observed around 20 GeV in the GRB spectrum when the baryon loading parameter is $\eta<\eta_{\star}/4$ which mainly occurs for a small number of shells (see Eq. (\ref{ETAS})).
At last, $pp$-collisions also generate a low energy emission of neutrinos because of the neutron decay which occurs after a proper neutron lifetime of $\sim$ 13 hours. In fact, the neutron decay produces antineutrinos with energy about 0.7 MeV in the co-moving frame. This leads to antineutrinos of 0.2 GeV for the observer.

\subsection{\textsf{Electron acceleration and energy limitation}}
\label{elec}

In this subsection, we analyse, the Fermi acceleration process concerning the electron population and we will use the same formalism than for the proton acceleration. Also, we assume that electrons and photons decouple at the photospheric radius $r_{\star}$. Thus, the electron acceleration starts at $r_{\star}$ if $r_{\star}>r_{b}$ and at $r_{b}$ otherwise: we can define a radius corresponding to the beginning of the acceleration stage, namely $r_{acc}=\max(r_{\star},r_{b})$. In the co-moving frame,
, the acceleration time for an electron with an energy $\epsilon$ will be such that:
\begin{equation}
t_{acc}\simeq 4.3\times 10^{-4}\left(\frac{\kappa_{0}}{10}\right)\left(\frac{\eta}{300}\right)^{2/3}\left(\frac{B(r_{b})}{10^{4}\,G}\right)^{-1/3}\left(\frac{\epsilon}{1\,MeV}\right)^{1/3}\left(\frac{r}{r_{b}}\right)^{\frac{2+\alpha}{3}}\,s.
\end{equation}
We deduce that there are two main energy limitations in the electron acceleration beyond $r_{acc}$. The first one is the synchrotron limitation which is:
\begin{eqnarray}
\epsilon_{syn} \simeq 6.5\times 10^{2}\,\left(\frac{\kappa_{0}}{10}\right)^{-3/4}\,\left(\frac{\eta}{300}\right)^{-1/2}\left(\frac{B(r_{b})}{10^{4}\,G}\right)^{-5/4}\left(\frac{r}{r_{b}}\right)^{(5\alpha -2)/4}\,MeV.
\label{eesyn}
\end{eqnarray}

\begin{figure}[!h]
\begin{center}
\includegraphics[totalheight=8cm]{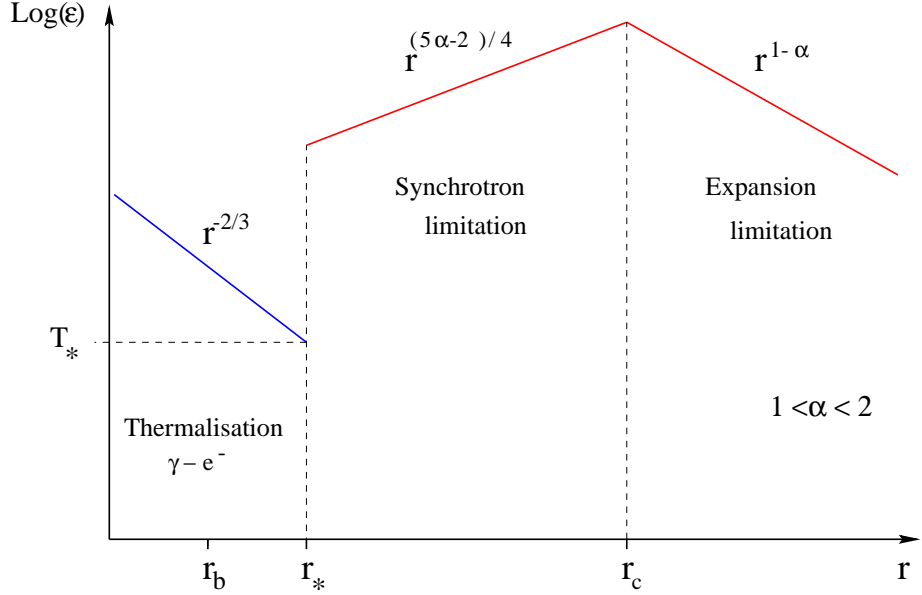}
\caption[10pts]{Diagram of energy limitation in the co-moving frame for the electrons. The position of the radius $r_{c}$ depends on the magnetic field index $\alpha$ and the electron energy at $r_{c}$, $\epsilon_{c}$ varies from $10^{5}$ to $10^{7}$ MeV.}
\label{elect}
\end{center}
\end{figure}

\noindent The second one is the expansion limitation:
\begin{eqnarray}
\epsilon_{exp} \simeq 1.2\times 10^{7}\,\left(\frac{\kappa_{0}}{10}\right)^{-3}\,\left(\frac{\eta}{300}\right)\,\left(\frac{B(r_{b})}{10^{4}\,G}\right)\left(\frac{r}{r_{b}}\right)^{1-\alpha}\,MeV.
\label{eeexp}
\end{eqnarray}
At $r_{b}$, the synchrotron limitation is the strongest one but we can define a radius, namely $r_{c}$,  where these two limitations are equal and beyond which the main limitation is the expansion one:
\begin{equation}
r_{c}=\left[ 5.4\times 10^{-5}\left(\frac{\kappa_{0}}{10}\right)^{9/4}\,\left(\frac{\eta}{300}\right)^{-3/2}\,\left(\frac{B(r_{b})}{10^{4}\,G}\right)^{-9/4}\right]^{\frac{4}{6-9\alpha}}\,r_{b}\,.
\end{equation}
For $\alpha =2$, $r_{c}\simeq 26\,r_{b}$, but for $\alpha$ close to 1, this radius is beyond $10^{5}\,r_{b}$. The electron energy increases and reaches the cut-off energy, $\epsilon_{c}$, at the radius $r_{c}$ (see Fig. \ref{elect}): this energy is very dependent on the magnetic field index $\alpha$ ($B\propto r^{-\alpha}$). We have:
\begin{equation}
\epsilon_{c}\simeq 6.5\times 10^{2}\,\left[ \left(\frac{\kappa_{0}}{10}\right)^{6\alpha-3}\left(\frac{\eta}{300}\right)^{\frac{2-5\alpha}{2}}\right]^{\frac{1}{2-3\alpha}}
\left[\left(5.4\times 10^{-5}\right)^{5\alpha-2}\left(\frac{B(r_{b})}{10^{4}\,G}\right)^{-3}\right]^{\frac{1}{6-9\alpha}}MeV.
\end{equation}
For $\alpha =2$, $\epsilon_{c}\simeq 4.5\times 10^{5}$ MeV and, for $\alpha$ close to 1, this energy reaches $1.2\times 10^{7}$ MeV.\\

\begin{figure}[!h]
\begin{center}
\includegraphics[totalheight=5.5cm]{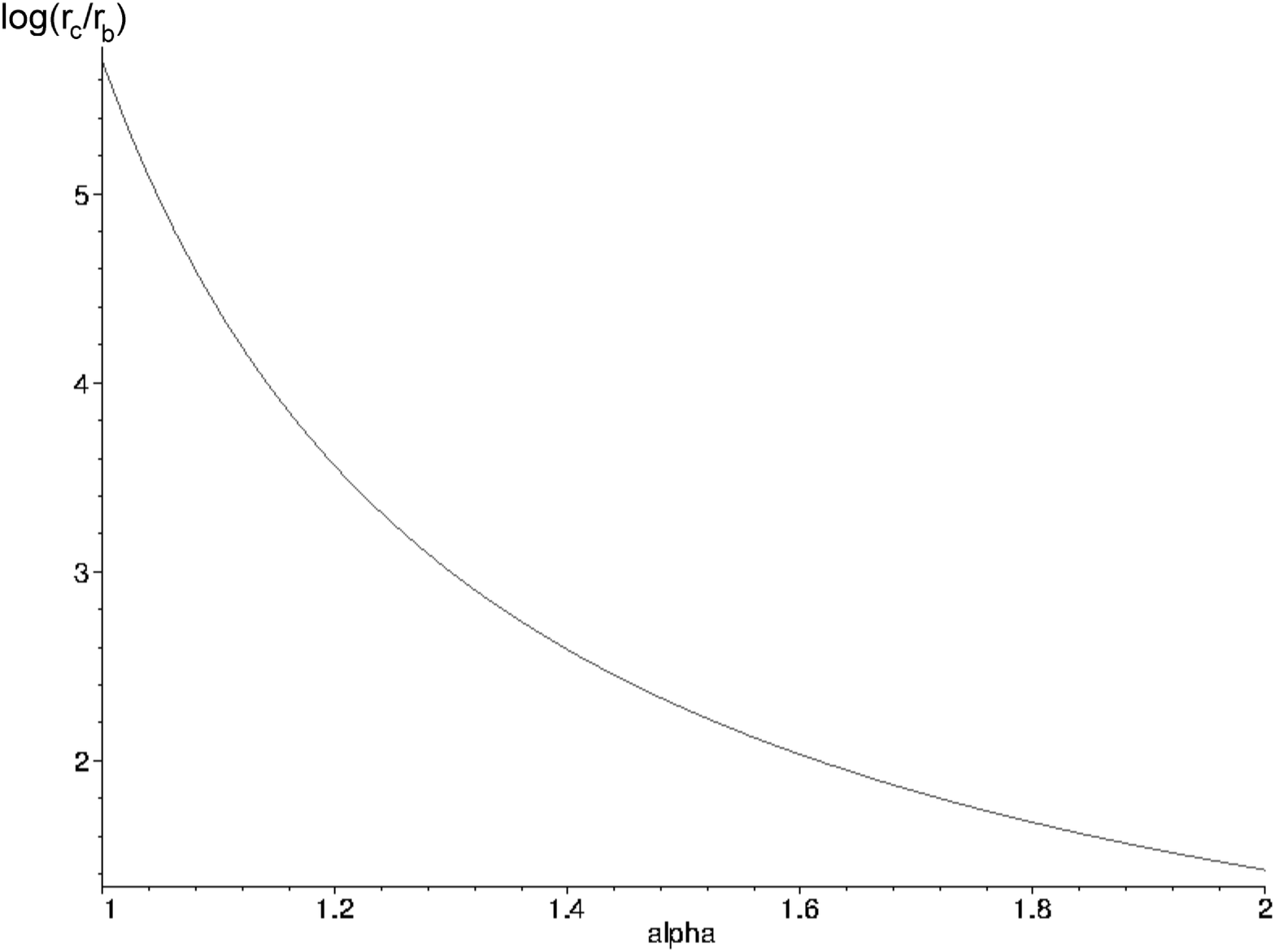}
\includegraphics[totalheight=5.5cm]{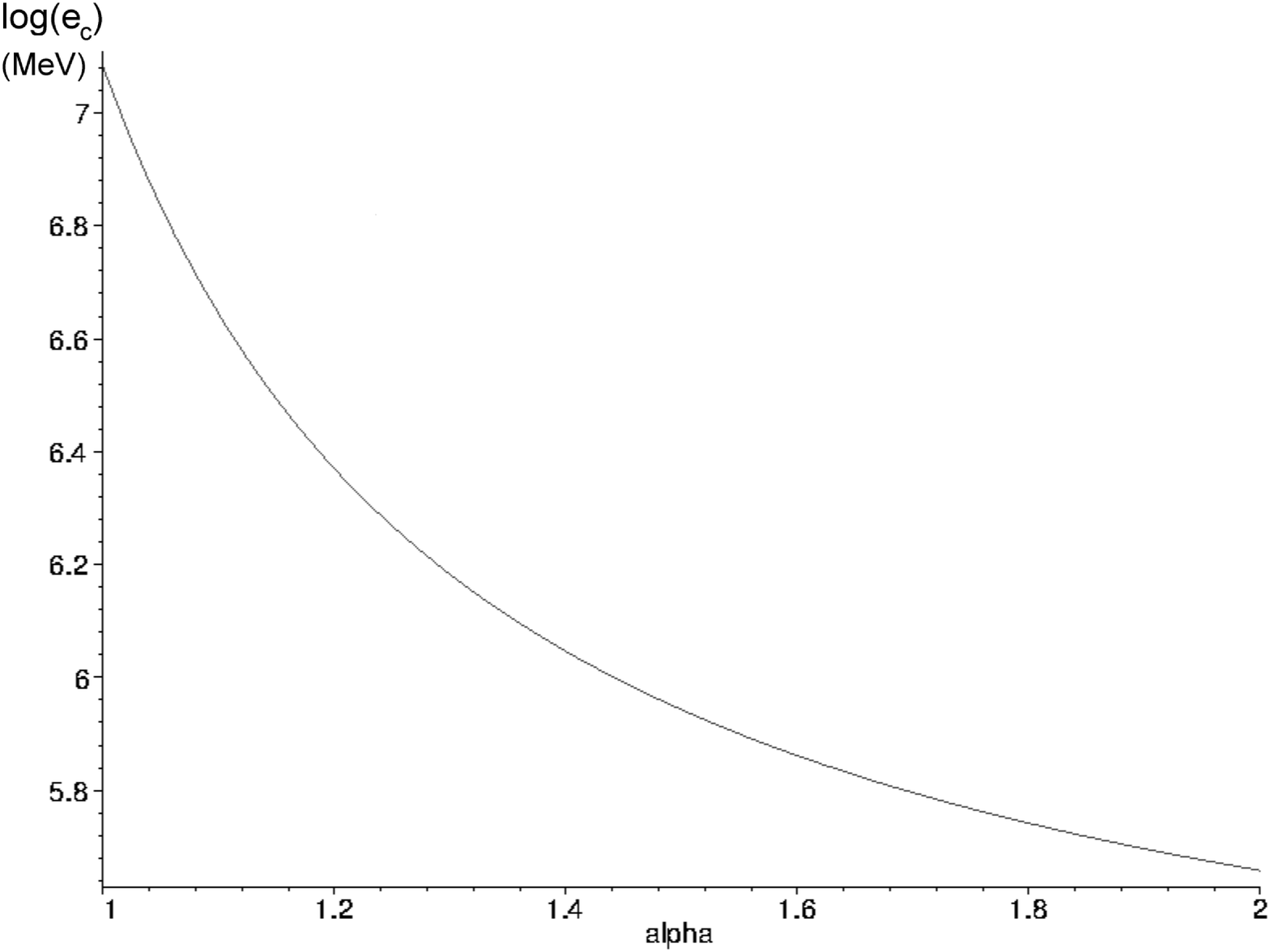}
\caption[10pts]{Left: Variation of $r_{c}/r_{b}$ as a function of the magnetic field index $\alpha$ for $\eta=300$. Right: Variation of the cut-off energy $\epsilon_{c}$.}
\label{therm2}
\end{center}
\end{figure}

\noindent For an electron with an energy $\epsilon$, the energy near which the synchrotron emission is a maximum can be expressed by:
\begin{equation}
h\,\nu_{syn}\simeq 7.0\times 10^{-4}\left(\frac{B(r_{b})}{10^{4}\,G}\right)\left(\frac{r}{r_{b}}\right)^{-\alpha}\left(\frac{\epsilon}{1\,MeV}\right)^{2}\,eV,
\label{esm}
\end{equation}
and, according Eq. (\ref{eesyn}), is of the order of 0.3 keV at $r_{b}$, in the co-moving frame, for accelerated electrons around the synchrotron energy limit.\\
Moreover, this energy reaches a maximum value at $r_{c}$ and we easily deduce that it is independent on $\alpha$: indeed, it can be written in the co-moving frame:
\begin{equation}
h\,\nu_{c}\simeq 2.0\times 10^{2}\left(\frac{\kappa_{0}}{10}\right)^{-3}\left(\frac{\eta}{300}\right)^{1/3}\,keV.
\label{egb}
\end{equation}
As we will see in the next section, this energy constitutes a cut-off energy in the high energy emission spectrum, and turns out to be remarkably independent on the magnetic field. Contrary to the case of proton acceleration, the additional acceleration by the scattering off the hydromagnetic fronts is not operating because the transit time of the electrons accross any shell is too long compared to the synchrotron loss time for an energy above the estimate given by Eq.(\ref{eesyn}).\\

\section{\textsf{Consequences on high energy emission}}
\label{HEE}

In this section, we focus on the high energy emission due to both proton and electron population and we consider a magnetic field strength that decreases like $r^{-2}$.

\subsection{\textsf{Synchrotron emission by electrons}}

We have seen in the previous section that electron acceleration is limited first by the synchrotron losses and, beyond $r_{c}$, by the expansion losses. We can define the electron energy limitation, namely $\epsilon_{b}$, by $\epsilon_{b}(r)=\epsilon_{syn}(r)$ up to $r_{c}$ and  $\epsilon_{b}(r)=\epsilon_{exp}(r)$ beyond.\\
According to Eqs. (\ref{esm}) and (\ref{egb}), there is a synchrotron emission over 3 energy decades between $0.3$ and $200$ keV in the co-moving frame, the highest energy corresponding to the cut-off energy $h\,\nu_{c}$. Assuming an electron density, namely $\rho(\epsilon,r)$, at the distance $r$ and for the energy $\epsilon$ such that
\begin{equation}
\rho(\epsilon,r)\propto \epsilon^{-2}\,\exp\left(-\epsilon/\epsilon_{b}(r)\right)\,,
\end{equation}
we deduce a local synchrotron spectrum i.e depending on the distance $r$ which can be written:
\begin{equation}
s_{e}(\nu,r)\propto \int \rho(\epsilon,r)\,P(\epsilon,r)\,\delta(\nu-\epsilon^{2}\,g(r))\,d\epsilon\,,
\label{se}
\end{equation}
where $P(\epsilon,r)\propto\epsilon^{2}\,g(r)^{2}$ is the total radiated power for an electron and defining $g(r)= \nu_{syn}\,\epsilon^{-2}$ i.e $g(r)\propto B(r)$ (see Eq. (\ref{esm})).
The integration of Eq. (\ref{se}) easily gives:
\begin{equation}
s_{e}(\nu,r)\propto g(r)\,\left(\frac{g(r)}{\nu}\right)^{1/2}\exp\left(-\sqrt{\frac{\nu}{g(r)}}\,\frac{1}{\epsilon_{b}(r)}  \right)\,.
\label{seb}
\end{equation}
Moreover, according to Eq. (\ref{esm}) we can define a local cut-off frequency, namely $\nu_{\star}$, by $\nu_{\star}(r)=g(r)\,\epsilon_{b}(r)^{2}$ so that we have $\nu_{\star}(r)\propto r^{2}$ up to $r_{c}$ and $\nu_{\star}(r)\propto r^{-4}$ beyond. Thus, the integration of Eq. (\ref{seb}) leads to a spectrum:
\begin{equation}
S_{e}(\nu)\propto \nu^{-1/2}\,\left[ \int_{r_{b}}^{r_{c}} r^{-3}\,\exp\left(-\sqrt{\frac{\nu}{\nu_{\star}(r_{c})}}\,\frac{r_{c}}{r} \right)\,dr+\int_{r_{c}}^{r_{d}} r^{-3}\,\exp\left(-\sqrt{\frac{\nu}{\nu_{\star}(r_{c})}}\,\frac{r^{2}}{r_{c}^{2}} \right)\,dr\right]\,,
\label{Se}
\end{equation}
where $\nu_{\star}(r_{c})$ corresponding to $\nu_{c}$ is the high cut-off frequency.\\
Considering the case $\nu < \nu_{c}$ i.e around a few 0.3 keV in the co-moving frame, we show in appendix that Eq.(\ref{Se}) could be simplified and we obtain:
\begin{equation}
S_{e}(\nu)\propto \nu^{-1}\,.
\end{equation}
Thus, electrons provide a gamma-ray emission which is constituted by a thermal component below a few 0.3 keV (see Sect. \ref{radpar}) and by a synchrotron component above. In the observer frame, the energy separating these two components could constitute the usual $E_{peak}$ (see e.g \citet{Band1993}) which is the energy for which the gamma-ray emission is the most important (i.e the peak energy of the burst). However most of the observations do not clearly exhibit such  $\nu^2$-spectrum in the low energy range. This $E_{peak}$ is determined by the lowest value of the synchrotron cut off as displayed by the limitation energy diagram, which corresponds to an emission at $r_b$. Moreover it depends on the baryon loading parameter $\eta$ because $E_{peak}\simeq \eta \,h\,\nu_{syn}(r_{b})$ and can vary from a few 30 keV to about 300 keV in the observer frame. The energy spectrum of the gamma-ray emission will be such that
\begin{eqnarray}
S(\nu)\propto \left\{ \begin{array}{ccl}
& \nu^{2} & \hspace{0.8cm} {\rm for}  \hspace{0.3cm} h\,\nu \lesssim E_{peak} \\
& \nu^{-1} & \hspace{0.8cm} {\rm for}
\hspace{0.3cm} E_{peak} \lesssim h\,\nu < \eta \, h\,\nu_{c}
\end{array} \right.\, .
\end{eqnarray}

With those estimates, there is a possible SSC generation under some conditions only: in fact, if $r_{acc}>r_{b}$, the electrons are quickly accelerated beyond 10 GeV and the SSC process, with keV photons, is in the Klein-Nishina regime ($\bar{\gamma}_{e}\,h\,\nu_{syn}\gg m_{e}\,c^{2}$ where $\bar{\gamma}_{e}$ is the average Lorentz factor of electrons). Because $\sigma_{KN}\ll \sigma_{T}$, this leads to an associated photospheric radius smaller than $r_{acc}$, so that there is no SSC generation. But, if $r_{acc}\simeq r_{b}$, because of a strong synchrotron energy limitation for electrons,  the SSC generation would give rise to a very low gamma emission in GeV range which would not be observable.\\

At this stage, it is useful to remark that if we would have used a Bohm scaling, the electron energy cut off due to synchrotron emission would be much higher:
\begin{equation}
\label{GBOH}
\gamma_e^{max} \simeq 3.2 \times 10^5 \left(\frac{\kappa_0}{10}\right)^{-1/2}\left(\frac{B(r_b)}{10^4\,G}\right)^{-1/2}\left(\frac{r}{r_b}\right) \, .
\end{equation}
This limit, that increases with distance contrarily to Kolmogorov one, would lead to a prohibitively high synchrotron spectrum.

\subsection{\textsf{Synchrotron emission by protons}}

Let us first consider the acceleration of protons at internal
shocks. The proton acceleration via the usual Fermi acceleration
is only limited by the expansion losses as we have seen in the
Sec. \ref{protacc}. The previous calculation for electrons can be
transposed to protons. For protons in the co-moving frame, the
synchrotron emission is maximum at the energy (for $B\propto
r^{-2}$):
\begin{equation}
h\,\nu_{syn}\simeq 1.0\times 10^{-6}\left(\frac{B(r_{b})}{10^{5}\,G}\right)\left(\frac{r}{r_{b}}\right)^{-2}\left(\frac{\epsilon}{1\,GeV}\right)^{2}\,eV.
\label{esmp}
\end{equation}
In the observer frame, the resulting synchrotron spectrum for protons will be such that $S_{p}(\nu)\propto \nu^{-1}$.
 According to Eq. (\ref{EEXP}), this spectrum extends from $10^{-7}$ to 1 eV in the co-moving frame.
  These low energy photons can interact with the accelerated electrons via the inverse Compton scattering:
  in this case for which $\bar{\gamma}_{e}\,h\,\nu\ll m_{e}\,c^{2}$ with a high average Lorentz factor
 ($\bar{\gamma}_{e}\geq 10^{3}$) of the electrons, the cross section is
 $\sigma_{c}\simeq \bar{\gamma}_{e}^{2}\,\sigma_{T}$. The associated photospheric radius, namely $r_{ph}$,
 is such that (see Eq. (\ref{TAUS})): $r_{ph}\simeq\bar{\gamma}_{e}\,r_{\star}\gg r_{b}$.
 During the internal shock phase, the inverse Compton effect thus produces photons with an energy amplified
 by a factor $\bar{\gamma}_{e}^{2}$ and which have an energy spectrum ($\propto \epsilon^{-1/2}$) extending
 from $0.1$ to over $10^{6}$ eV in the co-moving frame i.e from $10$ eV to $0.1$ GeV for the observer.  \\

Considering now UHECRs generated by the additionnal Fermi process
with Lorentz factors, namely $\gamma$, in the range
$[10^{8},10^{9}]$ in the co-moving frame \citep{DGGP04}: according
to Eq. (\ref{UHECR2}), the synchrotron energy spectrum will be such
that $S_{uhecr}(\nu)\propto \nu^{-1/2}$ and, following Eq.
(\ref{esmp}), UHECRs radiate synchrotron photons with an energy
that scales like $r^{-2}$ because of the magnetic field
decreasing. Moreover, for a magnetic field decreasing like $r^{-2}$, the minimal value at $r_{b}$ must be $\geq 10^{5}$ G. According to the acceleration process, we have shown
 that a sizeable fraction of the UHECR component is not generated before a distance $r_{cr}$ of a
few tens of $r_{b}$ typically between $50\,r_{b}$ and $100\,r_{b}$ and no longer after a few $r_{cr}$. This UHECR generation radius $r_{cr}$
 will determine the synchrotron emission range of UHECRs. In fact, because we have
\begin{equation}
\epsilon_{syn}^{p}=h\,\nu^{uhecr}_{syn}\simeq 10\,\left( \frac{B(r_{b})}{10^{5}\,G}
\right)\,\left( \frac{r}{r_{b}} \right)^{-2}\,\left(
\frac{\gamma}{10^{8}} \right)^{2}\,GeV\,, \label{euhe}
\end{equation}
the photons produced by UHECRs may extend from 1 MeV to 400 MeV in the co-moving frame.\\

We have to examine, now, the consequences of a pair creation process between these hadronic photons and the electronic ones. We have previously seen that the energy of the electronic photons reaches a maximum value of 200 keV at the distance $r_{c}$ that can be write
\begin{equation}
r_{c}\simeq 148 \left(\frac{B(r_{b})}{10^{5}\,G}\right)^{3/4}\,r_{b}\,,
\end{equation}   
for $B\propto r^{-2}$ so that we have $r_{cr}<r_{c}$ if $r_{cr}\simeq 100\,r_{b}$. According to the previous results (see Sect. \ref{elec}) and the Eq. (\ref{euhe}), the pair creation process happens if the product
\begin{equation}
\epsilon_{syn}^{p}\,\epsilon_{syn}^{e}\simeq
70\,\left(\frac{B(r_{b})}{10^5\,G}\right)^{2}\left(\frac{r}{r_{b}}\right)^{-4}\left(\frac{\epsilon_{b}}{1\,MeV}\right)^{2}\,\left(
\frac{\gamma}{10^{8}} \right)^{2}\,(keV)^{2}\label{equdist}\,,
\end{equation}
is higher than $2\,(m_{e}\,c^2)^2$ i.e $\simeq 5.2\times 10^5$ (keV)$^2$, where $\epsilon_{syn}^{e}$ is the energy of electronic photons. The threshold Lorentz factor, $\gamma_{th}$, beyond which the hadronic photons undergo a pair creation can be defined by
\begin{eqnarray}\label{gmax}
\gamma_{s}=\left\{ \begin{array}{cc} 2.3\times
10^{8}\left(\frac{B(r_{b})}{10^5\,G}\right)^{1/4}\, \hspace{0.2cm}
 \rm{for}\,\hspace{0.2cm}\,r_{cr}\leq r \leq r_{c}\,,\\
2.3\times
10^{8}\left(\frac{B(r_{b})}{10^5\,G}\right)^{-2}\,\left(\frac{r}{r_{c}}\right)^{3}
\hspace{0.2cm}
 \rm{for}\,\hspace{0.2cm}\,r > r_{c}\,.
\end{array}\right.
\end{eqnarray}
Thus, hadronic photons for which the Lorentz factor is between $10^{8}$ and $\gamma_{s}$ never undergo the pair creation process: so, the corresponding energy for hadronic photons will be in the range $[\epsilon_{syn}^{p}(\gamma=10^{8}), \epsilon_{syn}^{p}(\gamma_{th})]$ such that
\begin{equation}
\epsilon_{syn}^{p}(\gamma=10^{8})\simeq
1.0\,\left(\frac{B(r_{b})}{10^5\,G}\right)\left(\frac{r}{r_{cr}}\right)^{-2}
MeV\,,
\end{equation}
and,
\begin{equation}
\epsilon_{syn}^{p}(\gamma_{th})\simeq
5.4\,\left(\frac{B(r_{b})}{10^5\,G}\right)^{3/2}\,\left(\frac{r}{r_{cr}}\right)^{-2}
MeV\,,
\end{equation}
with $r_{cr}\simeq 100\,r_{b}$ and for $r_{cr}\leq r \leq r_{c}$.
Beyond $r_{c}$, the cut-off energy for hadronic photons will be
\begin{equation}
\epsilon_{syn}^{p}(\gamma_{th})\simeq
2.5\,\left(\frac{B(r_{b})}{10^5\,G}\right)^{-9/2}\,\left(\frac{r}{r_{c}}\right)^{4}
MeV\,.
\end{equation}

Therefore, the previous estimates allow to predict that an observer can detect synchrotron photons emitted by UHE-protons from a few 0.1 GeV to a few 10 GeV. We can remark that, if $r_{cr}>225\,r_{b}$ then, $\epsilon_{syn}^{p}(\gamma=10^{8})<200$ keV, and this emission may not be observable because of the electronic synchrotron component. \\

To end this review of all the possible absorption effects, we note that the interaction of these photons with themselves, which is kinematically possible, leads to a negligible opacity, namely, $\tau_{\gamma \gamma}\sim (E_{syn}/E)\,\tau_{\star}\sim (10^{-5}-10^{-4})\,\tau_{\star}$.

Let us estimate the corresponding global radiated energy that we
will compare to the energy in the UHECR component. For an
$E^{-2}$-spectrum, assuming an uniform flux during $\Delta t_w$,
the energy $E_{\star}$ in the CR-component above $\gamma_{0} \geq
1$ is
\begin{equation}
\label{ESTAR0}
E_{\star} = \Gamma\, m_{p}\, c^{3} \, \Omega \,r^{2} \int_{\gamma_{0}}^{\gamma_{max}} \rho(\gamma) \, \gamma \,\ud\gamma \, \Delta t_{w} \ ,
\end{equation}
where $\rho(\gamma) = n_{\star} \, \gamma_{0} \, \gamma^{-2}$ from $\gamma_{0}$ up to $\gamma_{max}\simeq 10^{9}$ and the number of cosmic rays above $\gamma_{0}$, $N_{\star}= n_{\star}(r_{b})\,\Omega\,r_{b}^{2}\,c\,\Delta t_{w}$. We obtain
\begin{equation}
\label{ESTAR} E_{\star} = \Gamma \, m_{p} \,c^{2}
\,N_{\star}\,\gamma_{0}
\,\log\left(\frac{\gamma_{max}}{\gamma_{0}}\right)\,.
\end{equation}
This can be simply compared to the GRB energy, $E$, since the total energy injected in protons,
 $E_{p}=E_{\star}(\gamma_{0}=1)$, is a sizeable fraction of $E$. We can write
\begin{equation}
\label{ESSE}
\frac{E_{\star}}{E} \sim 1-\frac{\log \gamma_{0}}{\log \gamma_{max}}  \, ,
\end{equation}
because $N_{p}\simeq \gamma_{0}\,N_{\star}$.
Concerning the UHECRs with $\gamma_{0}=10^{8}$, we reasonably find that $E_{uhecr}/E\simeq 10^{-1}$.
 Let us come back now to the estimate of the radiated energy which could be observed.
 Beyond the generation radius, $r_{cr}$, each proton synchrotron radiates a total energy
\begin{equation}
\label{IESYN} e^{syn}_{p}(\gamma) =
\frac{4}{3}\,\left(\frac{m_e}{m_p}\right)^2\, \sigma_T\, \Gamma \,c\,\gamma^2
\int_{r_{cr}}^{r_d} \frac{B(r)^{2}}{8\pi}\,\ud t \, ,
\end{equation}
which leads to:
\begin{equation}\label{IESYNb}
e^{syn}_{p}(\gamma) \simeq
9.6\times 10^{7}\,\left(\frac{\eta}{300}\right)^{3}\,\left(\frac{B(r_{b})}{10^{5}\,G}\right)^{2}\,
\left(\frac{\gamma}{10^{8}}\right)^{2}\,\left(\frac{r_{b}}{r_{cr}}\right)^{3}\,ergs.
\end{equation}
Thus, because $\ud N_{uhecr}/\ud \gamma \simeq
10^{8}\,N_{uhecr}\,\gamma^{-2}\simeq N_{p}\,\gamma^{-2}$ where
$N_{uhecr}$ ($\gamma \geq 10^{8}$) and $N_{p}$ are respectively
the total number of UHECRs and protons, the global emission which
can be observed will have an energy:
\begin{equation}\label{GESYNO}
E^{syn}_{uhecr} \simeq
\int_{10^{8}}^{10^{9}}N_{p}\,e^{syn}_{p}(\gamma)\,\gamma^{-2}\,\ud
\gamma\,,
\end{equation}
and, with $N_{p}\simeq E/(\eta\,m_{p}\,c^{2})$, we obtain:
\begin{equation}\label{GESYNOb}
\frac{E^{syn}_{uhecr}}{E} \simeq 19\,\left(\frac{\eta}{300}\right)^{2}\,\left(\frac{B(r_{b})}{10^{5}\,G}\right)^{2}\,
\left(\frac{r_{b}}{r_{cr}}\right)^{3}\,.
\end{equation}
For $r_{cr}=50$-$100\,r_{b}$, this ratio is between $10^{-5}$ and $10^{-4}$. This leads to a number of photons of about $10^{-3}$-$10^{-5}$ cm$^{-2}$ for a GRB located at 1 Gpc and pointing towards observatory. But, if we consider a magnetic field strength slightly higher of $10^{5}$ G at $r_{b}$, this number will increase around $10^{-3}$ cm$^{-2}$. A few GeV-photons may be detectable by the GLAST instrument which will have an effective area of about $10^{4}$ cm$^{2}$. At last, this number of detected photons could increase up to about one hundred for a GRB located at 100 Mpc.\\
This will constitute a very interesting signature of the UHECR generation in GRBs and will provide
us some constraints on the internal shock model. The only possible competitive emission in that range could be produced by SSC process at the reverse shock and/or the early external shock under exceptional conditions \citep{GranotGuetta03}.

\section{\textsf{Conclusion}}

The combined analysis of both electron and proton acceleration together with the observation data leads to interesting conclusions in the frame of conservative assumption about the magnetic field and the scattering of particles off its irregularities. Indeed with a sub-equipartition magnetic field, that decreases like $1/r^2$, but concentrated in the shells invoked to account for the light curve, we have shown the following points.
\begin{itemize}
  \item The low gamma ray spectrum is satisfactorily explained by the synchrotron radiation of the electrons that are accelerated at the internal shocks with a spectrum displaying the expected peak emission. The lowest energy part of the spectrum would be likely explained by a thermal component.
  \item The possibility of an UHECR generation under the previously stated conditions, as proposed in \citet{DGGP04}, is confirmed and the estimated flux is in agreement with the expected one \citep{Waxman95} to account for the Cosmic Ray spectrum around the ankle. A detailed estimate of the GRB contribution of the UHECRs to the Cosmic Ray spectrum, which is currently recorded by the Pierre Auger Observatory, has been proposed by \citet{BahcallWaxman01}; it displays the expected excess around the GZK threshold.
  \item The generation of cosmic rays in GRBs often starts before the fireball becomes transparent to pp-colisions. This gives rise to a gamma emission around 20 GeV due to $\pi^0$-decay. This emission is not contaminated by the SSC-emission of the electrons, because that latter emission is in Klein-Nishina regime.
We can also expect that neutrons are produced by these pp-collisions, they decay after some travel and thus generate a significant neutrino flux of 200 MeV (observer).
\item The most plausible signature of UHECR generation is not related to the p$\gamma$-process, but more likely to their synchrotron emission. Indeed, the ratio of the corresponding luminosities $L_{p\gamma}/L_{syn} \simeq U_s /U_{mag}$, where $U_s$ is the energy density of the soft photons and $U_{mag}$ the energy density of the magnetic field, turns out to be of order of unity. Since the energy of the neutrinos ($\sim$TeV for the observer) emitted through the p$\gamma$-process is much higher than the energy of the gamma-photons ($\sim$GeV) emitted by the synchrotron process, the neutrinos are thousand times less numerous than the photons. Thanks to our conservative assumption about the magnetic field, a gamma-ray flux can be so emitted and could be observed. Indeed we found a natural range of gamma energies for which the fireball is transparent to pair creation process, because of the chronology of the emissions during the fireball expansion and of the decrease of the magnetic field in $r^{-2}$. This range typically extends from a few hundreds of MeV to 10 GeV. It turns out that the observation of GeV photons should be a signature of UHECRs with probably no confusion. This could be observed by GLAST. At 1 Gpc, a $10^5$ G field would lead to a significant number of events which, of course, increases at shorter distances. This would be a very interesting signature of the UHECR generation in GRBs.
\end{itemize}

\appendix

\section{Appendix}

We detail the exact determination of the synchrotron spectrum as
defined by Eq. (\ref{Se}). Let us rewrite this expression:
\begin{equation}
S_{e}(\nu)\propto \nu^{-1/2} \left[ \int_{r_{b}}^{r_{c}}\frac{1}{r^{3}}\,\exp(-\alpha\,\frac{r_{c}}{r})\,\ud r + \int_{r_{c}}^{r_{d}}\frac{1}{r^{3}}\,\exp(-\alpha\,\frac{r^{2}}{r_{c}^2})\,\ud r \right]\,,
\end{equation}
where $\alpha \equiv \sqrt{\nu/\nu_{c}} \geq \sqrt{\nu_{syn}(r_{b})/\nu_{c}} \simeq 3.9 \times 10^{-2}$.\\

\noindent First, a simple integration by parts leads to
\begin{equation}
\int_{r_{b}}^{r_{c}}\frac{1}{r^{3}}\,\exp(-\alpha\,\frac{r_{c}}{r})\,\ud r = \frac{1}{r_{c}^{2}\alpha^{2}}\,\left[(\alpha+1)\,\exp(-\alpha)-\frac{\alpha+\lambda}{\lambda}\,\exp(-\alpha/\lambda)\right]\,,
\end{equation}
with $\lambda=r_{b}/r_{c}\simeq 1/26$ for $B\propto r^{-2}$.\\

\noindent For the second integral, because $r_{c}\ll r_{d}$, we can assume $r_{d}\rightarrow +\infty$ and, next a first integration by parts, we can write for instance:
\begin{equation}
\int_{r_{c}}^{r_{d}}\frac{1}{r^{3}}\,\exp(-\alpha\,\frac{r^{2}}{r_{c}^2})\,\ud r = \frac{r_{c}^{2}}{2\alpha}\,\left(\frac{\exp(-\alpha)}{r_{c}^{4}}-4\,\int_{r_{c}}^{+\infty}\frac{1}{r^{5}}\,\exp(-\alpha\,\frac{r^{2}}{r_{c}^2})\,\ud r \right)
\end{equation}
An iterative integration by parts gives, with some manipulations, the following result:
\begin{equation}
\int_{r_{c}}^{r_{d}}\frac{1}{r^{3}}\,\exp(-\alpha\,\frac{r^{2}}{r_{c}^2})\,\ud r = \frac{\exp(-\alpha)}{2\,r_{c}^{2}\,\alpha}\,\left( \sum_{n=0}^{+\infty}\frac{(-1)^{n}(n+1)!}{\alpha^{n}}\right),
\end{equation}
where the sum of the serie is convergent whatever $\alpha$ according to the generalized hypergeometric function (or Barnes's extended hypergeometric function). For $\alpha=\alpha_{min}\simeq 3.9\times 10^{-2}$, this sum is equal to $3.5\times 10^{-2}$ and for $\alpha=1$, the sum reaches 0.4.\\
Thus, the synchrotron spectrum $S_{e}(\nu)$ is such that:
\begin{eqnarray}
S_{e}(\nu) \propto \nu^{-1} \,\left[\left(\frac{\alpha+1}{\alpha}+\frac{1}{2}\sum_{n=0}^{+\infty}\frac{(-1)^{n}(n+1)!}{\alpha^{n}}\right)\exp(-\alpha)-\frac{\alpha+\lambda}{\lambda\,\alpha}\,\exp(-\alpha/\lambda)\right]\,.
\end{eqnarray}
For $\alpha_{min}\leq \alpha < 1$, the expression between brackets is quasi constant so that  $S_{e}(\nu)\propto \nu^{-1}$. Beyond $\alpha =1$, the same expression leads to a spectrum decreasing like $\nu^{-s}$ with $s\in [1.5,2]$.

\bibliographystyle{aa}
\bibliography{/gagax1/ur1/dgialis/THESE/BIBFILES/biblio}

\end{document}